\documentclass{amsart}
\usepackage[utf8]{inputenc}

\usepackage[normalem]{ulem} 
\usepackage[super]{nth} 
\usepackage{setspace}
\usepackage{silence}
\usepackage{amsmath, amsthm,amssymb}
\usepackage{float}
\usepackage{bm}
\usepackage{natbib}
\usepackage{graphicx}
\usepackage{pgfplots}
\usepackage{pgfplotstable}
\pgfplotsset{compat = newest}

\renewcommand{\le}{\leqslant}
\renewcommand{\ge}{\geqslant}

\newcommand{\R}{\mathbb{R}}

\begin{document}

\title{Pulsatile Annular Flow with Coaxial Fluid Jet}

\author{Jean-Luc Boulnois }

\address{FineHeart, Pessac 33600 , France, jlboulnois@msn.com}


\date{August 2025}


\maketitle


\begin{abstract}
This study provides exact analytical solutions for both steady-state and pulsatile annular flows in coaxial cylindrical systems. It also examines  the effects of a synchronized inner tube high velocity jet and its potential impact on annular blood flow. The presence of such a fluid jet significantly enhances the velocity profile and flow rate across the annular section. These models offer valuable insights into optimizing flow performance in potential cardiovascular applications. 
\end{abstract}

\section{Introduction}

This study focuses on  annular flows of viscous fluids between coaxial pipes under both steady-state and oscillatory conditions.
\newline

Section 2 presents the governing equations of $\textit{unsteady annular flows}$ between two rigid  coaxial  cylindrical  tubes, including appropriate boundary conditions. The analysis of periodic flow generated by pulsatile  pressure gradients draws on Womersley's work on arterial blood flow \cite{Womersley}. Understanding the interplay between inertial and viscous forces  is central to analyzing flow dynamics in this  geometry.
\newline

Section 3 reviews $\textit{steady-state annular flow}$ solutions  and examines the effect of a  constant velocity fluid jet  ejected  from the inner tube. Exact analytic solutions establish the enhancing jet's impact on the velocity profile and  flow rate.
\newline

In Section 4,  $\textit{oscillatory annular flow}$  solutions are presented assuming the same geometry. The effect of a  $\textit{pulsatile fluid jet}$ synchronized with flow oscillations is  analyzed. Both  exact analytical and asymptotic solutions are derived.
\newline

Section 5 extends the analysis to the cardiovascular context where a pulsatile blood jet interacts with the native heart flow in a  coaxial geometry. Future work will address the more complex and realistic scenarios involving elastic heart walls.

\section{Pulsatile Annular Flow Between  Straight Coaxial Tubes}

The study of axial flows of viscous fluids in tubes is governed by the continuity and Navier-Stokes equations. The present study assumes  the fluid to be homogeneous, incompressible with constant density $\rho$, and Newtonian with constant dynamic viscosity $\mu$. 
\newline

A one-dimensional axisymmetric flow is assumed to occur between two coaxial, straight, rigid, smooth cylindrical tubes that are  sufficiently long for the flow to be fully developed, i.e. with no radial or azimuthal velocity component. Given the moderate velocities considered, the flow is assumed to be laminar.  The pressure gradient  driving the flow varies periodically with time, and gravity effects  are considered negligible.
\newline

Let $r$ be the radial coordinate and  $u(r,t)$ the longitudinal flow velocity parallel to the positive $x$-axis at radius $r$ and time $t$.  The continuity equation together with the Navier-Stokes  equation written in cylindrical coordinates  are respectively 
\begin{equation} \label{eq:1}
\frac{\partial u}{\partial x} = 0
\end{equation}

and
\begin{equation} \label{eq:2}
\rho \frac{\partial u}{\partial t} = -\frac{\partial p}{\partial x} + \mu \left(\frac{\partial^2 u}{\partial r^2} + \frac{1}{r}\frac{\partial u}{\partial r}\right) 
\end{equation}

Equation (\ref{eq:2}) expresses the instantaneous balance between inertial and viscous forces driven by an  oscillating axial pressure gradient. It must be solved subject to the "\textit{no-slip}" boundary condition, i.e. zero velocity respectively on the outer surface of the inner cylindrical tube of radius $R_1$ and on the inner surface of the outer coaxial tube of radius $R_2$
\begin{equation} \label{eq:3}
u(R_1, t) = u(R_2, t) = 0
\end{equation}

The pressure is only function of $x$ and $t$ and is independent of $r$, i.e. ${\partial p} / {\partial r = 0}$.   Following Womersley and Uchida  \cite{Womersley}, \cite{Uchida}, since Eq. (\ref{eq:2}) is linear the periodic  pressure gradient driving the  flow in the $x$-direction over a distance $L$  may be decomposed into a Fourier series where $\omega$ is the natural angular frequency of the first harmonic component $(n = 1)$ and $\widetilde{P}_n$ is the   $n^{th}$-harmonic complex amplitude  (divided by $L$)
\begin{equation} \label{eq:4}
\frac{\partial p}{\partial x} = \frac{d p}{d x} =  \sum_{n=0}^{N}{\widetilde{P}_n e^{i n \omega t}}
\end{equation}

The  flow velocity field is  similarly decomposed into a Fourier series in which $u_n(r)$, function of $r$ alone, represents the amplitude of each harmonic velocity contribution
\begin{equation} \label{eq:5}
u(r,t) = \sum_{n=0}^{N}{u_n (r) e^{i n \omega t}}
\end{equation}

To quantify the balance between  inertial and viscous flow components,  the Navier-Stokes  equation is scaled by introducing  a scaled radius $\xi$  of order unity, based on a generic  radius $R$ 
defined as $r = \xi R$, and a scaled  time $\theta$, defined  as $\theta = \omega t$  
\begin{equation} \label{eq:6}
\frac{\partial u}{\partial \theta} = -\frac{1}{\rho \omega}\frac{\partial p}{\partial x} + \frac{1}{\alpha^2} \left(\frac{\partial^2 u}{\partial \xi^2} + \frac{1}{\xi}\frac{\partial u}{\partial \xi}\right) 
\end{equation}

Using the kinematic viscosity $\nu = \mu / \rho$,  this scaled Navier-Stokes equation introduces the dimensionless Womersley number $\alpha$ defined as \cite{Womersley}
\begin{equation} \label{eq:7}
\alpha = R \sqrt{\frac{\omega}{\nu}}
\end{equation}

The   Womersley number $\alpha $ measures the relative magnitudes of inertial to viscous flow contributions. When $\alpha $ is small,  inertial  effects are negligible relative to  viscous effects, and the Navier-Stokes velocity solutions exhibit standard    Poiseuille-like flow profiles that vary in magnitude but not in shape (see Section 3). 
\newline

As shown in (\ref{eq:6}), a high Womersley number indicates that inertial effects   dominate over  viscous effects, resulting in nearly flat  velocity profiles in the flow core. The Navier-Stokes equation then simplifies to
\begin{equation} \label{eq:8}
\frac{d u}{d t}  = - \frac{1}{\rho} \sum_{n=0}^{N}{\widetilde{P}_n } e^{i n \omega t}
\end{equation}

The flat core velocity profile, solution of  (\ref{eq:8}),  oscillates periodically like plane waves in inviscid flows. To satisfy the $\textit{no-slip}$ condition (\ref{eq:3}), a thin viscous boundary layer   develops in the vicinity of the  tube walls, matching the core velocity to the zero  velocity at the surfaces (see Section 4). 
\newline

Section 5  considers flows  in the  high unsteady regime in which the respective Womersley numbers associated with each concentric cylindrical surface satisfy
\begin{equation} \label{eq:9}
\alpha_1 \simeq 5 \le  r\sqrt{ \frac{\omega}{\nu}}  \le \alpha_2 \simeq 20
\end{equation}

\section{Steady-State Annular Flow  With   Fluid Jet}

In the  steady-state the velocity is independent of time (${\partial u} / {\partial t} = 0$) and the pressure gradient between the tube extremities  is  constant: it is labeled $- \widetilde{P}$,  and, for $n\ge 1$, all $\widetilde{P}_n = 0$ and  all $u_n = 0$. 
\newline

From the equation of motion (\ref{eq:2}), the  governing  $2^{nd}$ order ordinary differential equation (ODE)  becomes
\begin{equation} \label{eq:10}
\mu \left(\frac{d^2 u}{d r^2} + \frac{1}{r}\frac{d u}{d r}\right) = \frac{d p}{d x}
\end{equation}

The steady-state annular flow velocity component $u_0(r)$  for  $n = 0$  within the concentric cylindrical geometry is obtained by solving  (\ref{eq:10}) subject to the \textit{no-slip} boundary condition (\ref{eq:3}).  The solution is simplified by introducing several dimensionless parameters. First, the ratio of the respective tube radii is defined as $\lambda = \frac{R_2}{R_1}$  with  $\lambda \ge 1$. Next, the  radial coordinate $r$ is rendered dimensionless by using the conventional ratio $\hat{r} = \frac{r}{R_2}$, implying $\frac{1}{\lambda} \le \hat{r} \le 1$. The basic Poiseuille flow in a rigid cylindrical tube of radius $R_2$   corresponds to $R_1 = 0$, i.e. $\lambda \to \infty$. 
\newline

Further, a dimensionless parameter $\beta$ related to the \textit{no-slip} condition at $\hat{r} = 1/\lambda$ is defined as
\begin{equation} \label{eq:11}
\beta = \frac{1- \frac{R_1^{2}}{R_2^{2}}}{\ln \frac{R_2}{R_1}} = \frac{1 - \frac{1}{\lambda^2}}{\ln \lambda}
\end{equation}

Noticeably $\beta = 2$ when $\lambda = 1$ (i.e. no flow), while $\beta = 0$ when $\lambda \to \infty$. Lastly, all velocities are measured in units of the peak Poiseuille velocity $u_p = \frac{\widetilde{P} R_2^{2}}{4 \mu}$.
\newline

Applying the \textit{no-slip} boundary conditions $u(\frac{1}{\lambda}) = u(1) = 0$  to (\ref{eq:10}) yields the  following known steady-state annular flow solution \cite{Landau}
\begin{equation} \label{eq:12}
u_0(\hat{r}) = u_p \left( 1 - \hat{r}^2 + \beta \ln(\hat{r}) \right)
\end{equation}

Poiseuille's standard steady-state parabolic velocity solution is obtained  by setting $R_1 = 0$, i.e. $\beta = 0$ in  (\ref{eq:12}). In this equation, the negative logarithmic contribution $\beta \ln(\hat{r})$ results from applying the $1^{st}$ boundary condition $u_0(\frac{1}{\lambda}) = 0$, whereas the $2^{nd}$ boundary condition ensures that $u_0(1) =0$. The velocity  reaches a maximum at an intermediary neutral surface of radius $\hat{r}_m = \sqrt{\frac{\beta}{2}}$. Consequently, the corresponding annular peak velocity $u(\hat{r}_m)$ is  considerably reduced relative to $u_p$: for example, at $\lambda = 4$, $\hat{r}_m = 0.581$ and $u(\hat{r}_m)$ = 0.295$u_p$.
\newline

The annular mass flow rate $Q$  is obtained by integrating the velocity field over the  cross-section 
\begin{equation} \label{eq:13}
Q = 2 \pi \rho \int_{R_1}^{R_2} r u(r) dr
\end{equation}

Inserting the velocity solution (\ref{eq:12}) into (\ref{eq:13})  yields the steady-state annular flow rate $Q_0$ 
\begin{equation} \label{eq:14}
Q_0 = Q_p \left(1 - \frac{1}{\lambda^2} \right) \left(1 + \frac{1}{\lambda^2} - \beta \right)
\end{equation}

where $Q_p = \frac{\pi}{8} \frac{\widetilde{P} R_2^{4}}{\nu}$ is the standard $R^4$-dependent Poiseuille flow rate which is recovered from (\ref{eq:14}) by letting $\lambda \to \infty$. Relative to a Poiseuille flow, two factors contribute to reducing the flow rate in this geometry: the $1^{st}$ is the annular  cross-sectional reduction, which rapidly diminishes with increasing $\lambda$; the $2^{nd}$ factor is associated with the inner  surface $\textit{no-slip}$ boundary condition. For instance, at $\lambda = 4$, $Q_0$ is about 36 \% of $Q_p$.
\newline

\underline{3.1   Annular Flow with Coaxial Fluid Jet}
\newline 

Coaxial flows are commonly used  in various engineering and industrial applications for mixing fluids, chemicals, or particles. Most studies generally focus on free shear flows where mixing is not confined by an outer pipe.
\newline

Numerous theoretical and experimental studies have investigated the turbulent mixing of coaxial jet flows inside cylindrical conduits \cite{Mikhail}, \cite{Buresti} or with large velocity ratio between  jet flow and  outer stream \cite{Rehab}.
\newline

We  consider  a  high velocity jet flow  with a constant velocity $V_0$ of order  $u_p$ and a plug-like velocity profile ejected from a small hollow  tube of radius $R_1$ confined inside a larger concentric tube of radius $R_2$. Ejection takes place in the normal plane  $x = 0$,   entrance to the  mixing region $x \ge 0$ with the outer annular flow.  
\newline

Although at high jet velocities turbulence may arise in this region   (see Section 5.2), in the following both the steady-state annular  and  jet streams are treated as laminar with identical density,  viscosity, and temperature. In the near field $x \ge 0$, the fluid jet is assumed to continuously maintain its  velocity: in this simple model the applicable boundary conditions for the  induced secondary stream become
\begin{equation}\label{eq:15}
u(R_1) = V_0  \quad     \text{and}  \quad u(R_2) = 0
\end{equation}

\begin{figure}[h!]
\centering
\includegraphics[scale=0.35]{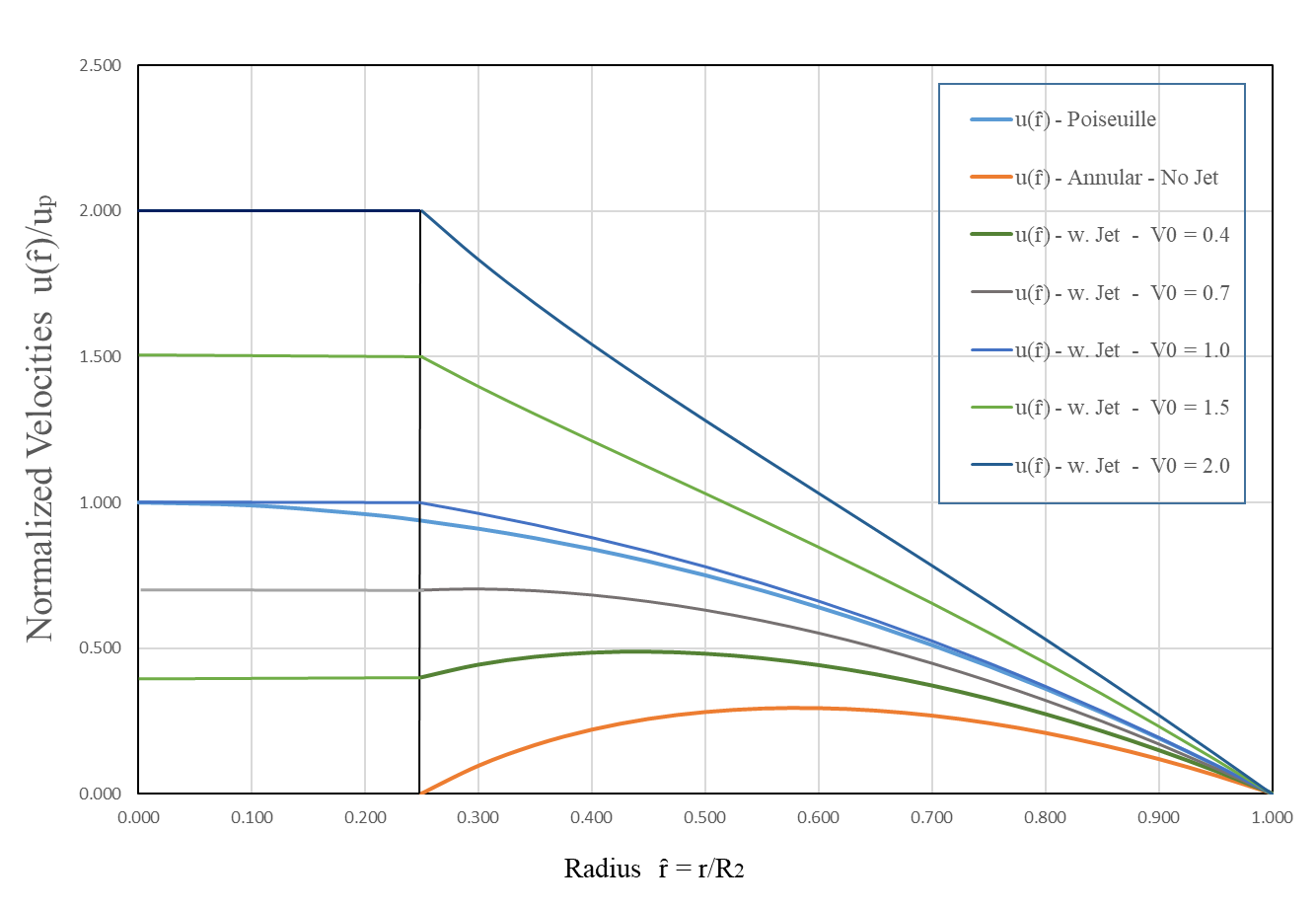}
\caption{Near-field steady-state  annular velocity profiles  $u(\hat{r})/u_p$ as a function of $\hat{r}$ for $\lambda = 4$: without jet (\ref{eq:12}), and with jet (\ref{eq:16}) at velocities $\overline{V}_0 = 0.4, 0.7, 1.0, 1.5$ and $2.0$; Poiseuille profile for reference. }
\label{fig:Fig1}
\end{figure}

Solving (\ref{eq:10}) with these new boundary conditions, and introducing the dimensionless jet velocity $ \overline{V}_0 = \frac{V_0}{u_p}$, yields the modified annular steady-state velocity solution, labeled $u_{0,j} (\hat{r})$
\begin{equation} \label{eq:16}
u_{0,j}(\hat{r}) = u_p \left( 1 - \hat{r}^2 + \left( \beta - \frac{\overline{V}_0}{\ln \lambda} \right) \ln(\hat{r}) \right)
\end{equation}

The annular velocity profile $u_{0,j}(\hat{r})$ includes a jet-induced positive contribution $u_p \frac{\overline{V}_0}{\ln \lambda} \ln( \frac{1}{\hat{r}}) $ superposed to  solution (\ref{eq:12}). Figure 1 displays the normalized  annular flow $u_{0,j}(\hat{r})/u_p$ resulting from the near-field mixing with the constant velocity fluid jet.  The parameters are $\lambda = 4$ and $\overline{V}_0 = 0.4, \; 0.7, \; 1.0, \; 1.5$ and $2.0$, respectively. Also shown is the no-jet steady-state annular velocity profile $u_0(\hat{r})/u_p$. For reference the corresponding Poiseuille parabolic velocity profile is also indicated. 
\newline

The presence of high velocity fluid jet  exiting the inner tube at $x = 0$   significantly enhances the  velocity profile by $\textit{entraining}$ the annular flow across the entire  cross-section. The jet overcomes the $\textit{no-slip}$ constraint  by increasing the velocity at the $ \hat{r} = 1/\lambda$ interface.  When $\overline{V}_0 \le 1  $ the jet increases the annular peak velocity $u(\hat{r}_{m,j}) $ while displacing the corresponding peak radius  $\hat{r}_{m,j} = \sqrt {\frac{1}{2}(\beta- \frac{\overline{V}_0}{ \ln\lambda})}$ closer to the inner tube than in the standard case (\ref{eq:12}). When $\overline{V}_0 \ge 1 $ the high velocity jet entrains the annular flow further with velocities greater than the Poiseuille velocity.

\begin{figure}[h!]
\centering
\includegraphics[scale=0.35]{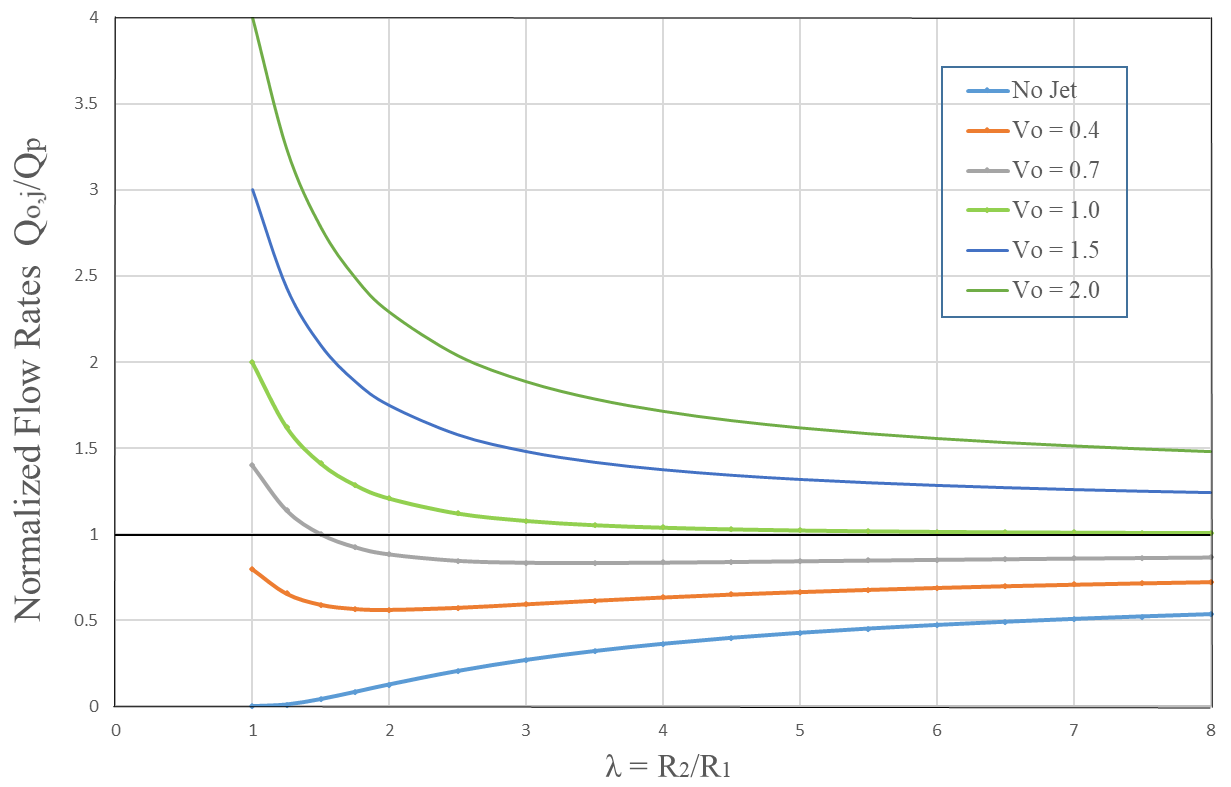}
\caption{Normalized flow rates  $Q_{0,j}/Q_p$ as a function of $\lambda = R_2/R_1$: without jet, and with jet at $\overline{V}_0 = 0.4, 0.7, 1.0, 1.5$, and 2.0. }
\label{fig:Fig2}
\end{figure}

This jet-induced annular flow entrainment  significantly increases the total flow rate.
By integrating  (\ref{eq:13}) together with the velocity solution (\ref{eq:16}), the total annular flow rate becomes
\begin{equation} \label{eq:18}
Q_{0,j} =  Q_0 + Q_p \beta \overline{V}_0 = Q_0 + \beta \frac{\lambda^2}{2} Q_{jet}
\end{equation}

Since the flow is assumed to be fully developed, relative to the jet-free flow rate  (\ref{eq:14}), the presence of the central  fluid jet of flow rate $Q_{jet} = \pi R_1^2 \rho V_0$ provides an \textit{additional positive flow contribution} of magnitude $ \beta \frac{\lambda^2}{2} Q_{jet}$  over the entire annular cross-section with an amplifying factor $ \beta \frac{\lambda^2}{2} \ge 1$. 
\newline  

This fluid jet entrainment inside the annulus gap is  demonstrated on Figure 2 which displays the normalized flow rate $Q_{0,j}/Q_p$ as a function of the ratio $\lambda $. The graph compares the  "no jet" flow rate case (\ref{eq:14}) to five jet cases  with respective  ejections velocities $\overline{V}_0 = 0.4, \; 0.7,  \; 1.0, \; 1.5$  and 2.0. The line $Q_{0,j}/Q_p = 1$ corresponds to the Poiseuille flow rate.
\newline

 Regardless of the value of $\lambda$, the jet flow increases the total annular flow rate. When  $\lambda \simeq 1^{+}$  and $\overline{V}_0 \le 1.0$, the gap between the coaxial tubes is very small therby reducing the flow rate, while near the inner tube surface   the  jet velocity is insufficient to  fully compensate   the $\textit{no-slip}$ constraint of the no-jet case resulting in a minimum for the enhanced  flow rate. However, beyond this point the  multiplying factor $Q_{0,j}/Q_0$ is significant: for example for $\lambda = 4$ and $\overline{V}_0 = 0.4$  the  factor is approximately 1.75. For $\overline{V}_0 \ge 1.0$ there is no minimum:  the annular flow rate exceeds the Poiseuille  threshold and the enhancement factor is substantial. For example for $\lambda = 4$ and $\overline{V}_0 = 2.0$, the  factor reaches 4.74. 
\newline

Solution (\ref{eq:18})   demonstrates the \textit{amplifying effect} of high velocity jet-flow entrainment on the  steady-state annular flow rate. This property was leveraged in a recent mechanical cardiac assist device  \cite{Cima1}, and in an intra-aortic circulatory device \cite{Heuring}. 
\newline

\section{Oscillatory Annular Flow: Solutions of the Equations of Motion}

We  consider the case of an annular flow driven by  an axial periodic pressure gradient  with a resulting periodic pulsatile velocity profile.
\newline 

By introducing equations (\ref{eq:4}) and (\ref{eq:5}) into the equation of motion (\ref{eq:2}) the partial differential equation for each harmonic component of the velocity field becomes a linear $2^{nd}$ order ODE
\begin{equation} \label{eq:19}
\nu \left(\frac{d^2 u_n}{d r^2} + \frac{1}{r}\frac{d u_n}{d r}\right) - i n \omega u_n = \frac{\widetilde{P}_n}{\rho}
\end{equation}

The steady-state velocity profile $u_0(\hat{r})$ corresponds to the $n = 0$ harmonic,  (\ref{eq:12}).  The general solution of (\ref{eq:19}) for $n \ge 1$ admits the  particular solution (inertial flow component (\ref{eq:8}))
\begin{equation} \label{eq:20}
u_n^{(part)} (r) = i \frac{\widetilde{P}_n}{n \rho \omega}
\end{equation}

Making the following change of variables into (\ref{eq:19}) and  introducing the function $y(x)$ representing the $n^{th}$-harmonic velocity amplitude, we define
\begin{equation} \label{eq:21}
 u_n = \frac{\nu}{\omega} y(x)   \quad     \text{and }  \; r =x \sqrt{\frac{\nu}{\omega}}  
\end{equation}

The resulting equation becomes a well-known $2^{nd}$ order Bessel ODE
\begin{equation} \label{eq:22}
x^2 \frac{d^2 y}{dx^2} + x \frac{dy}{dx} -i n x^2 y = x^2 \frac{\widetilde{P}_n}{\mu}
\end{equation}

Solutions of this ODE can either be obtained through Bessel functions of zero$^{th}$ order (Section 4.1) or equivalently  through Kelvin functions (Section 4.2).
\newline

\underline{4.1   Bessel Functions  Solutions }
\newline

In Eq. (\ref{eq:22}), a last change of variables, $x = n^{-1/2} i^{-3/2} z$, results in the canonical form of Bessel's equation of zero$^{th}$ order  whose  homogeneous solutions  are expressed in terms of the respective Bessel functions of the $1^{st}$ kind  $J_0(z)$ and  $2^{nd}$ kind  $Y_0(z)$ 
\begin{equation} \label{eq:23}
z^2 \frac{d^2 y}{dz^2} + z \frac{dy}{dz} + z^2 y = 0
\end{equation}

\underline{4.1.1  Pulsatile Velocity Solutions}
\newline

The general solution of the non-homogeneous linear ODE  (\ref{eq:19}) for the $n^{th}$-harmonic velocity amplitude, in which $\beta_n = \sqrt{n}i^{3/2}$, is 
\begin{equation} \label{eq:24}
u_n(r) =  i \frac{\widetilde{P}_n}{n \rho \omega} + C_n J_0\left(\beta_n r \sqrt \frac{\omega}{\nu}\right) + D_n Y_0\left(\beta_n r \sqrt \frac{\omega}{\nu}\right) 
\end{equation}

The arbitrary constants $C_n$ and $D_n$ are determined from the  \textit{no-slip} boundary conditions at $r = R_1$ and $r = R_2$, yielding  a system of two linear equations based on two Womersley  numbers, respectively $\alpha_1 = R_1 \sqrt{ \frac{\omega}{\nu}}$ and  $\alpha_2 = R_2 \sqrt{ \frac{\omega}{\nu}}$
\begin{equation} \label{eq:25}
 \begin{array}{l}
      {C_n  J_0(\beta_n \alpha_1) + D_n Y_0 (\beta_n \alpha_1) = - i \frac{\widetilde{P}_n}{n \rho \omega}  } \\
      {C_n  J_0 (\beta_n \alpha_2) + D_n Y_0 (\beta_n \alpha_2) = - i \frac{\widetilde{P}_n}{n \rho \omega}  } 
 \end{array}
\end{equation} 

Introducing the determinant $\Delta_n(\alpha_1, \alpha_2)$
\begin{equation} \label{eq:26}
\Delta_n(\alpha_1, \alpha_2) = J_0(\beta_n \alpha_1) Y_0 (\beta_n \alpha_2) - J_0 (\beta_n \alpha_2) Y_0 (\beta_n \alpha_1)
\end{equation}

the coefficients $C_n$ and $D_n$, which only depend on $\alpha_1$ and $\alpha_2$, then become
\begin{equation} \label{eq:27}
 \begin{array}{l}
{C_n = -  i \frac{\widetilde{P}_n}{n \rho \omega} \left( \frac{Y_0 (\beta_n \alpha_2) - Y_0 (\beta_n \alpha_1)}{\Delta_n(\alpha_1, \alpha_2)}   \right)  }  \\
{D_n = +  i \frac{\widetilde{P}_n}{n \rho \omega} \left( \frac{J_0 (\beta_n \alpha_2) - J_0 (\beta_n \alpha_1)}{\Delta_n(\alpha_1, \alpha_2)}   \right)  }
\end{array}
\end{equation}

The final exact analytical solution of Eq. (\ref{eq:19}) for the amplitude of the  $n^{th}$-harmonic oscillatory  velocity component  is
\begin{multline} \label{eq:28}
u_n(r) = i \frac{\widetilde{P}_n}{n \rho \omega} \left( 1 - \frac{Y_0 (\beta_n \alpha_2) - Y_0 (\beta_n \alpha_1)}{\Delta_n(\alpha_1, \alpha_2)} J_0\left(\beta_n r \sqrt{\frac{\omega}{\nu}} \right) \right. \\
\left. + \frac{J_0 (\beta_n \alpha_2) - J_0 (\beta_n \alpha_1)}{\Delta_n(\alpha_1, \alpha_2)}  Y_0\left(\beta_n r \sqrt{\frac{\omega}{\nu}}\right) \right)
\end{multline}

 In  (\ref{eq:28}) all Bessel functions  contain the parameter $\beta_n = \sqrt{n} i^{3/2} $ which vanishes as $n \to\ 0$. Their  lower limits are evaluated  by recalling  the respective limiting forms   of the Bessel functions of the $1^{st}$ and $2^{nd}$ kind \cite{Abramowitz1} for small argument $z$ 
\begin{equation} \label{eq:29}
\begin{array}{l}
     J_0(z) = 1- \frac{z^2}{4} + O(z^4)    \\
     Y_0(z) = \frac{2}{\pi} (\ln{\frac{z}{2}} + \gamma) + O(z^2)
\end{array}
\end{equation}

where $\gamma$ is the Euler-Mascheroni constant and $Y_0(z)$ exhibits  a logarithmic singularity at $z = 0$. Inserting these limits into solution (\ref{eq:28}),  and recalling that $\hat{r} = r/R_2$,   yields
\begin{equation} \label{eq:30}
u_0(\hat{r}) = \left(i \frac{\widetilde{P}_n}{n \rho \omega}\right)\left(i n \frac{\alpha_2^2}{4}\right) \frac{1-\hat{r}^2 + \beta \ln(\hat{r})}{1+i n \frac{\alpha_2^2}{4}(1 - \beta(\ln (\frac{\beta_n \alpha_2}{2}) + \gamma))}
\end{equation}

Rearranging the first two factors and recalling that $\widetilde{P}_0 = - \widetilde{P}$, precisely yields the Poiseuille peak velocity $u_p $; as $n \to 0$,   the denominator of (\ref{eq:30}) tends to unity, and the steady-state solution (\ref{eq:12}) is recovered.
\newline

Remarkably, as $n \to 0$, the Bessel solution $J_0(\beta_n \alpha_2 \hat{r})$ of the $1^{st}$ kind in  (\ref{eq:28}) directly corresponds to the steady-state Poiseuille velocity contribution in (\ref{eq:12}),  whereas the logarithmic velocity contribution $\beta \ln(\hat{r})$  is derived from the Bessel solution $Y_0(\beta_n \alpha_2 \hat{r})$ of the $2^{nd}$ kind.
\newline

At any given time $t$, the entire pulsatile flow profile  is  obtained by  inserting the steady-state solution $u_0(r)$,   summing over all pulse harmonic amplitudes ($n \ge 1)$),   and taking the real part of the resulting complex function 
\begin{equation} \label{eq:31}
u(r,t) = u_0(r) + \Re \left(  \sum_{n=1}^{N}{u_n (r) e^{i n \omega t}}  \right)
\end{equation}

Notice that, due to the presence of the imaginary "i" in (\ref{eq:28}), the instantaneous velocity field $u(r, t)$ is out of phase by $\pi/2$ with the pressure wave.
\newline

The oscillatory flow amplitude in a straight cylinder with no inner coaxial tube ($R_1 = 0$), is directly obtained from   (\ref{eq:28}). Dividing numerator and denominator of each fraction in (\ref{eq:27}) by $Y_0 (\beta_n \alpha_1)$ and letting $\alpha_1 \to 0$ while using the limits of the zero$^{th}$ order Bessel functions  (\ref{eq:29}) yields
\begin{equation} \label{eq:33}
C_n = - i \frac{\widetilde{P}_n}{n \rho \omega} \frac{1}{J_0 (\beta_n \alpha_2)}
\end{equation}

Also, as $\alpha_1 \to 0$, due to the singularity of the Bessel function of the $2^{nd}$ kind $Y_0(\beta_n \alpha_1$), all coefficients  $D_n$   vanish. Therefore, in this $\alpha_1 \to  0$ limit (i.e. large  $\alpha_2/\alpha_1$ ratio),  the  $n^{th}$-harmonic oscillatory  velocity  amplitude   becomes
\begin{equation} \label{eq:35}
u_n (\hat{r}) \simeq i \frac{\widetilde{P}_n}{n \rho \omega} \left(1 -  \frac{J_0(\beta_n \alpha_2 \hat{r} ) }{J_0 (\beta_n \alpha_2)} \right )
\end{equation}

This is exactly the  velocity  solution obtained for $n = 1$ by Womersley for a pulsatile viscous flow inside a cylindrical tube of radius $R_2$,  \cite{Womersley}.
\newline 

The interplay between harmonics and viscous layer dynamics provide insights into the nature of pulsatile flows under different Womersley number regimes. In the  high Womersley number case, the inequalities  (\ref{eq:9}) have significant consequences for  expressing the solution  (\ref{eq:28}) in more manageable terms   since  asymptotic expansions of  Bessel or Kelvin functions can be used (see Section 4.2.1).
\newline

\underline{4.1.2   Pulsatile Annular Flow Rate}
\newline

Proceeding  as in  Section 3, the instantaneous pulsatile annular flow rate is obtained by  integration of (\ref{eq:13})  using  the  $n^{th}$-harmonic oscillatory velocity component (\ref{eq:28}), prior to summing over all harmonics. Standard integrals involving products like $z J_0(z)$ and $z Y_0(z)$ are respectively evaluated in terms of Bessel functions of the $1^{st}$ and $2^{nd}$ kind, but of the first order, namely $J_1(z)$ and $Y_1(z)$, \cite{Watson}. 
\newline

Expressing  the $n^{th}$-harmonic complex amplitude $\widetilde{P_n}$  in terms of its modulus and phase as $|\widetilde{P_n}| e^{i \phi_n} $, and omitting intermediate calculations, the exact  pulsatile annular flow rate  is 
\begin{multline} \label{eq:36}
Q(t) = Q_p\left( \left (1 - \frac{1}{\lambda^2} \right)  \left( 1 + \frac{1}{\lambda^2} -\beta - \frac{8}{\alpha_2^2} \sum_{n=1}^{N}{\frac{|\widetilde{P_n}|}{\widetilde{P}}\frac{\sin{(n \omega t + \phi_n)}}{n}} \right) + \right. \\
\frac{64}{\pi \alpha_2^4}  \Re  \sum_{n=1}^{N}{ \frac{|\widetilde{P_n}|}{\widetilde{P}} \frac{1}{n^2} \frac{e^{i (n \omega t + \phi_n)}}{\Delta_n(\alpha_1, \alpha_2)} }  +  \\
\frac{16}{\alpha_2^4} \Re \sum_{n=1}^{N}{ \frac{1}{n^{3/2}} \frac{|\widetilde{P_n}|}{\widetilde{P}} \frac{\alpha_2 (Y_0(\beta_n \alpha_1) J_1(\beta_n \alpha_2) - J_0(\beta_n \alpha_1) Y_1(\beta_n \alpha_2) ) }{\Delta_n(\alpha_1, \alpha_2) } e^{i(n \omega t + \phi_n - \frac{\pi}{4})}} +  \\
\left. \frac{16}{\alpha_2^4} \Re \sum_{n=1}^{N}{ \frac{1}{n^{3/2}}\frac{|\widetilde{P_n}|}{\widetilde{P}} \frac{\alpha_1 (J_1(\beta_n \alpha_1) Y_0(\beta_n \alpha_2) - Y_1(\beta_n \alpha_1) J_0(\beta_n \alpha_2) ) }{\Delta_n(\alpha_1, \alpha_2) } e^{i(n \omega t + \phi_n - \frac{\pi}{4})}}  \right)
\end{multline}

 The instantaneous pulsatile annular mass flow rate $Q(t)$ includes the mean annular  flow rate $Q_0$ from (\ref{eq:14}) corrected by a harmonic inertial flow contribution out of phase with the oscillating pressure. The additional  viscous flow components of order $O(\alpha_2 ^{-4})$ are negligible when $\alpha_2 \gg 1$. 
\newline

\underline{4.2   Pulsatile Annular Flow: Kelvin Functions Solutions}
\newline

The conventional separation of solution (\ref{eq:28}) into real and imaginary parts  yields cumbersome formulae  due to the presence of products of Bessel functions of complex argument $ \sqrt{n} i^{3/2 } \alpha$. In our case where the Womersley numbers  are large, use of the Kelvin functions renders the velocity amplitude solution more manageable.
\newline

By introducing the real variable $\xi = xn^{1/2}$, Eq. (\ref{eq:22}) becomes
\begin{equation} \label{eq:41}
\xi^2 \frac{d^2 y}{d\xi^2} + \xi \frac{dy}{d\xi} -i  \xi^2 y = \xi^2 (\frac{\widetilde{P}_n}{n \mu})
\end{equation}

The homogeneous part of this ODE  admits two independent solutions in terms of Kelvin functions of zero$^{th}$ order, respectively labeled $y_b(\xi) = ber(\xi)$ + $i$ $bei(\xi)$ and $y_k(\xi) = ker(\xi)$ + $i$ $kei(\xi)$, where $\xi \in$ $\R$ is  non-negative. The functions   $ber(\xi)$ and $bei(\xi)$ are the respective real and imaginary parts of the Bessel functions of  the $1^{st}$ kind; similarly, $ker(\xi)$ and $kei(\xi)$  pertain to the  the modified Bessel functions of the $2^{nd}$ kind \cite{Abramowitz1}.
\newline

The above change of variables implies $\xi = r \sqrt{ \frac{n\omega }{\nu}}$;  two boundary values $\xi_1 = \alpha_1 \sqrt{n}$ and $\xi_2 = \alpha_2 \sqrt{n}$ are introduced in terms of their respective Womersley numbers $\alpha_1$ and $\alpha_2$, with $\xi_1 \le \xi \le \xi_2$. ODE (\ref{eq:41}) is solved subject to the $\textit{no-slip}$ conditions at $\xi_1$ and $\xi_2$, yielding a system of two linear equations for the coefficients C and D
\begin{equation} \label{eq:42}
 \begin{array}{l}
      {C  y_b(\xi_1) + D y_k (\xi_1) = - i \frac{\widetilde{P}_n}{n \mu}  } \\
      {C  y_b (\xi_2) + D y_b (\xi_2) = - i \frac{\widetilde{P}_n}{n \mu}  } 
 \end{array}
\end{equation} 

The determinant $\Delta (\xi_1, \xi_2)$ of this system is
\begin{equation} \label{eq:43}
\Delta (\xi_1, \xi_2) = y_b(\xi_1) y_k(\xi_2) - y_b(\xi_2) y_k(\xi_1)
\end{equation}

The exact  analytical solution of the non-homogeneous ODE (\ref{eq:41}) for the $n^{th}$-harmonic velocity amplitude  $y(\xi)$  which satisfies the boundary conditions (\ref{eq:3}) becomes
\begin{equation} \label{eq:44}
y(\xi) = i \frac{\widetilde{P}_n}{n \mu} \left( 1 - \frac{y_k(\xi_2) - y_k(\xi_1)}{\Delta ( \xi_1, \xi_2)}y_b(\xi) +  \frac{y_b(\xi_2) - y_b(\xi_1)}{\Delta ( \xi_1, \xi_2)}y_k(\xi)    \right)
\end{equation}

Up to the multiplying factor $\frac{\nu}{\omega}$ for $y(\xi)$, this expression is identical to  solution (\ref{eq:28}) for the $n^{th}$-harmonic annular velocity component, \textit{albeit} written in terms of Kelvin functions which admit simple asymptotic expansions.
\newline

\underline{4.2.1 Asymptotic Solutions}
\newline

Per  (\ref{eq:9}), since $\alpha_1$ and $\alpha_2$  are large, it is reasonable to use the asymptotic expansions of the Kelvin functions   for the $ber$-function \cite{Abramowitz1},  namely
\begin{equation}  \label{eq:45}
y_b(\xi) = \frac{e^{\frac{\xi}{\sqrt{2}}}}{\sqrt{2 \pi \xi}} e^{i( \frac{\xi}{\sqrt{2}} - \frac{\pi}{8})}
\end{equation}

and for the $ker$-function \cite{Abramowitz1}
\begin{equation}  \label{eq:46}
y_k(\xi) =\sqrt{ \frac {\pi}{2\xi}}e^{- \frac{\xi}{\sqrt{2}}} e^{- i (\frac{\xi}{\sqrt{2}} + \frac{\pi}{8})}
\end{equation}

Using these asymptotic Kelvin functions, the determinant (\ref{eq:43}) simplifies to
\begin{equation} \label{eq:47}
\Delta (\xi_1, \xi_2 ) = \frac{e^{-i\frac{\pi}{4}}}{{\sqrt{\xi_1 \xi_2}}} \sinh((\xi_1-\xi_2) e^{i \frac{\pi}{4}})
\end{equation}

Substitution of these asymptotic  functions into (\ref{eq:44})  yields the  asymptotic solution of the inhomogeneous Kelvin ODE (\ref{eq:41})
which \textit{satisfies} the \textit{no-slip} boundary conditions at $\xi = \xi_1$ and $\xi = \xi_2$
\begin{equation} \label{eq:49}
y(\xi) =  i \frac{\widetilde{P}_n}{n \mu}  \left( 1 - \left(  \frac{\sqrt{\frac{\xi_1}{\xi}}\sinh((\xi_2-\xi) e^{i \frac{\pi}{4}}) + \sqrt{\frac{\xi_2}{\xi}}\sinh((\xi-\xi_1) e^{i \frac{\pi}{4}})}{\sinh((\xi_2-\xi_1) e^{i \frac{\pi}{4}})}          \right)  \right)
\end{equation}


\underline{4.2.2   Real Function Asymptotic Solutions }
\newline

Solution (\ref{eq:49}) is further  expressed in terms of simple $\textit{real functions}$ by observing that for large arguments, $ \sinh(z) \simeq e^z/2$ for $z \gg 1$ (the \textit{exponential approximation}). 
\newline 

Accordingly, this approximation neglects the contribution from the $e^{-z}/2$ part of $\sinh(z)$. In our case,  contributions of order  $O(e^{-(\sqrt n (\alpha_2 - \alpha_1))/\sqrt{2}})$ can thus be neglected: using $\alpha_1 = 5$ and  $\alpha_2 = 20$  from (\ref{eq:9}) together  with $n = 1$, the neglected contribution is of order $O(\epsilon)$ with $\epsilon = e^{-15/ \sqrt{2}}\simeq 2.4 \; 10^{-5}$.
\newline

Applying this exponential approximation to   solution (\ref{eq:49}) yields
\begin{equation} \label{eq:50}
y(\xi, t) =  i \frac{\widetilde{P}_n}{n \mu}  \left( 1 - \sqrt{\frac{\xi_1}{\xi}} e^{-(\xi - \xi_1) e^{i \frac{\pi}{4}}} - \sqrt{\frac{\xi_2}{\xi}}  e^{-(\xi_2 - \xi) e^{i \frac{\pi}{4}}}    \right) e^{ i n \omega t}
\end{equation}

As a consequence of the above approximation, in  (\ref{eq:50})  it should be noted  that $y(\xi, t)$  does not strictly match the \textit{no-slip} boundary conditions respectively at $\xi = \xi_1$ or $\xi = \xi_2$, but  with a negligible mismatch of order  $O(\epsilon)$.
\newline

Reverting to physical coordinates with $\xi - \xi_1= \sqrt{n} \alpha_1 (\lambda \hat{r} - 1)$ and $\xi_2 - \xi= \sqrt{n} \alpha_2 (1 - \hat{r})$,  solution (\ref{eq:50}) gives the complex amplitude of the $n^{th}$-harmonic  pulsatile velocity
\begin{multline} \label{eq:51}
u_n(\hat{r}, t) = i \frac{\widetilde{P}_n}{n \rho \omega}  \left(1 - \frac{1}{\sqrt{\lambda \hat{r}}}  e^{-\sqrt{\frac{n}{2}}\alpha_1 (\lambda \hat{r} - 1) } e^{i ({n \omega t - \sqrt{\frac{n}{2}} \alpha_1 (\lambda \hat{r} - 1) })}  \right.  -  \\  
\left. \frac{1}{\sqrt{\hat{r}}}  e^{-\sqrt{\frac{n}{2}}\alpha_2 (1-\hat{r}) } e^{i ({n \omega t - \sqrt{\frac{n}{2}} \alpha_2 (1-\hat{r}) } )} \right)
\end{multline}

Taking the real part of this complex  amplitude and expressing $\widetilde{P_n}$ in terms of its modulus and phase provides an asymptotic solution for the  oscillatory  $n^{th}$-harmonic velocity amplitude in terms of  \textit{real functions}
\begin{multline} \label{eq:52}
u_n(\hat{r}, t) =  u_p \frac{4}{\alpha_2^2}  \frac{|\widetilde{P_n}|}{n \widetilde{P}}  \left( \frac{1}{\sqrt{\lambda \hat{r}}}  e^{-\sqrt{\frac{n}{2}}\alpha_1 (\lambda \hat{r} - 1) } \sin{ \left( {n \omega t + \phi_n - \sqrt{\frac{n}{2}} \alpha_1 (\lambda \hat{r} - 1) } \right)}  + \right.  \\  
\left. \frac{1}{\sqrt{\hat{r}}}  e^{-\sqrt{\frac{n}{2}}\alpha_2 (1-\hat{r}) } \sin{ \left({ n \omega t + \phi_n - \sqrt{\frac{n}{2}} \alpha_2 (1-\hat{r}) } \right) }  -  \sin{ ( n \omega t + \phi_n )} \right) 
\end{multline}

At any  time $t$, the entire oscillatory velocity profile  is  obtained by  inserting the steady-state solution $u_0(\hat{r})$ given by (\ref{eq:12}) and   summing over all above real pulse harmonic amplitudes ($n \ge 1$)      
\begin{equation} \label{eq:53}
u(\hat{r},t) = u_0(\hat{r}) +  \sum_{n=1}^{N}{u_n (\hat{r}, t) }
\end{equation}

This constitutes an asymptotic \textit{real function}  solution of the Navier-Stokes equation (\ref{eq:2}) satisfying the boundary conditions (\ref{eq:3}), up to an infinitesimally small quantity of $O(\epsilon)$. These solutions are illustrated in the numerical results  in Section 5.1.
\newline

\underline{4.2.3   Transverse Waves}
\newline

Equation (\ref{eq:52})  demonstrates the interplay between   viscous and inertial effects: these effects induce \textit{transverse waves} in the flow. 
\newline

For each oscillatory  $n^{th}$-harmonic velocity component, in the vicinity of the respective annular surfaces $\hat{r} = 1/\lambda$ and $\hat{r} = 1$, as $\hat{r}$ approaches the boundary limit,   solution (\ref{eq:52}) represents transverse waves in a viscous layer. The   velocity amplitude $u_n(\hat{r}, t)$ is directed along the x-axis, yet the respective waves propagate  perpendicularly along the physical radius $\hat{r}$, in opposite directions towards each surface.

\begin{enumerate}
\item \underline{Amplitude}

The major property of these transverse waves is their rapid amplitude decay  in the fluid core: near each annular surface the waves decay exponentially in a thin viscous layer ("\textit{Stokes} layer") of thickness $O ((\alpha_1 \sqrt{n/2})^{-1})$ and $O ((\alpha_2 \sqrt{n/2})^{-1})$, respectively; the larger the Womersley number the thinner the viscous layer. In the flow core, the transverse wave \textit{depth of penetration}  is     $\sqrt{\frac{2 \nu}{n \omega}}$: it decreases when the frequency $\omega$ or the harmonic number $n$ increase, and it increases with the  viscosity.
\newline

\item \underline{Phase}  

Near both surfaces, the oscillation phase varies rapidly across their respective viscous layers. For example near the surface $\hat{r} = 1$, the  $n^{th}$-harmonic velocity field  corresponds to a transverse wave propagating towards the  boundary with the rapidly decreasing wavenumber  $ \sqrt{\frac{n \omega}{2 \nu}} (1-\hat{r}) $ according to
\begin{multline} \label{eq:54}
u_n(\hat{r}, t) \simeq  -\frac{|\widetilde{P_n}|}{n \rho \omega} \Big( \sin{(n \omega t + \phi_n)}  -  \Big.  \\
\left. e^{-\sqrt{\frac{n}{2}}\alpha_2 (1-\hat{r}) } \sin{\left( {n \omega t + \phi_n - \sqrt{\frac{n}{2}} \alpha_2 (1-\hat{r}) } \right) }  \right)
\end{multline}

Beyond the Stokes layer, the core velocity profile becomes nearly flat and independent of the viscosity:   inertial forces dominate.  In (\ref{eq:54}) for example, for $ \hat{r} \ll 1$ and large $\alpha_2 $, the $2^{nd}$ term decays exponentially: the velocity thus  oscillates like an inviscid wave in harmonic motion of angular frequency $n\omega $.  

Referring to (\ref{eq:52}), in the fluid core, the  two exponential terms can be neglected yielding the total  harmonic velocity solution of (\ref{eq:8}), oscillating out of phase with the pressure gradient 
\begin{equation} \label{eq:55}
u(t) = -\frac{4}{\alpha_2^2} u_p \sum_{n=1}^{N}{\frac{|\widetilde{P_n}|}{\widetilde{P}}\frac{\sin{(n \omega t + \phi_n)}}{n}} 
\end{equation}


\item \underline{Shear Stress}

To calculate  the pulsatile  shear stress component along an annular   surface, for example $r = R_2$,  we note that this force is parallel to the x-axis with a magnitude equal to the following x-component of the viscous stress tensor
\begin{equation} \label{eq:56}
\sigma_{xr} = \mu \frac{\partial u}{\partial r}|_{r=R_2} 
\end{equation}

Using this definition, the $n^{th}$-harmonic shear stress is obtained by differentiating (\ref{eq:54}) with respect to $\hat{r}$ and evaluating the shear rate at $\hat{r} = 1$ 
\begin{equation} \label{eq:57}
\sigma_{xr,n} = |\widetilde{P_n}| \sqrt{ \frac{\nu}{n \omega} } \; \sin{(n \omega t + \phi_n+ \frac{\pi}{4})} 
\end{equation}

Comparing this result with Eq. (\ref{eq:54}) shows that the oscillatory velocity and the wall shear stress are out of phase by $\frac{\pi}{4}$. As a result, it is possible for the flow velocity direction to become reversed as shown in Section 5.1.
\newline

\end{enumerate}

\underline{4.2.4   Pulsatile Annular Flow with Fluid Jet: Kelvin Functions}
\newline

Using the same geometric and flow conditions as those of Section 3.1, we consider a coaxial, periodic, high velocity  fluid jet $V_0(t)$, with a plug-like velocity profile, \textit{synchronous} with the pulsatile annular flow, decomposed in Fourier series as 
\begin{equation} \label{eq:37}
V_0(t)= \sum_{n=0}^{N}{V_n  e^{i n \omega t}}
\end{equation}

We seek an exact general solution of the equation of motion (\ref{eq:41}) in the presence of this  jet. At the interface $\xi = \xi_1$ the jet modifies the boundary condition to $y(\xi_2) = \frac{\omega}{\nu}V_n$. The system (\ref{eq:42}) of two linear equations for the coefficients C and D  becomes
\begin{equation} \label{eq:58}
 \begin{array}{l}
      {C  y_b(\xi_1) + D y_k (\xi_1) = \frac{\omega}{\nu}V_n - i \frac{\widetilde{P}_n}{n \mu}  } \\
      {C  y_b (\xi_2) + D y_b (\xi_2) = - i \frac{\widetilde{P}_n}{n \mu}  } 
 \end{array}
\end{equation} 

The determinant of this system is given by (\ref{eq:43}). Since Eq. (\ref{eq:41}) is linear, the exact solution for the  jet-induced complex $n^{th}$-harmonic annular  velocity amplitude $y_j(\xi)$  satisfying the boundary conditions at $\xi_1$ and $\xi_2$  is obtained by superposition with the jet-free exact solution (\ref{eq:49})
\begin{equation} \label{eq:59}
y_j(\xi) = y(\xi) +\frac{\omega}{\nu} \frac{V_n}{\Delta (\xi_1, \xi_2)}  \left( y_k(\xi_2) y_b(\xi) - y_b(\xi_2) y_k(\xi) \right)
\end{equation}

Using  asymptotic expressions of the Kelvin functions, an asymptotic solution of (\ref{eq:59}) satisfying the boundary conditions at $\xi_1$ and $\xi_2$, is 
\begin{equation} \label{eq:60}
y_j(\xi) = y(\xi) +\frac{\omega}{\nu} V_n \sqrt{\frac{\xi_1}{\xi}} \frac{\sinh((\xi-\xi_2) e^{i \frac{\pi}{4}})}{\sinh((\xi_1-\xi_2) e^{i \frac{\pi}{4}})}
\end{equation}


The jet velocity component $V_n$ is expressed in terms of its modulus and phase according to $V_n = |V_n| e^{i \psi_n}$. Reverting to physical coordinates, while further applying the exponential approximation to (53) since the Womersley numbers under consideration are large,   the asymptotic synchronous, pulsatile, instantaneous, jet-induced $n^{th}$-harmonic  annular velocity amplitude is expressed in terms of \textit{real functions}
\begin{equation} \label{eq:61}
u_{n,j}(\hat{r}, t) = u_n(\hat{r}, t) + |V_n| \frac{ e^{-\sqrt{\frac{n}{2}}\alpha_1 (\lambda \hat{r} - 1) }  }{\sqrt{\lambda \hat{r}}}  \cos{ \left( {n \omega t + \psi_n - \sqrt{\frac{n}{2}} \alpha_1 (\lambda \hat{r} - 1) } \right)} 
\end{equation}

Although this asymptotic approximation matches the boundary condition at $\hat{r} = 1/\lambda$, as expected, it exhibits a negligible mismatch of $O(\epsilon)$ at $\hat{r} = 1$.
\newline

The $n^{th}$-harmonic jet-induced flow-rate is obtained by integrating (\ref{eq:13}) together with the asymptotic solution (54): the exact solution of this integral is expressed in terms of incomplete Gamma functions  evaluated at the boundaries of  the normalized range $[1, \lambda]$. In this  range,  the exponential term $e^{-\sqrt{\frac{n}{2}}\alpha_1 (\lambda \hat{r} - 1) }$ dominates the  second term thereby simplifying the integration. Taking the real part of the resulting integral,  summing over all harmonics, and defining $|\overline{V}_n| = |V_n|/u_p $, yields an approximate asymptotic jet-induced annular flow rate $Q_j(t) $ as
\begin{equation} \label{eq:62}
Q_j(t) \simeq Q_p \frac{4}{\lambda^2 \alpha_1}  \sum_{n=1}^{N}{\frac{|\overline{V}_n|}{\sqrt{n}} \cos{(n \omega t + \psi_n - \frac{\pi}{4})}}  
\end{equation}

The enhancing effect of such a synchronous coaxial pulsatile jet on an oscillatory annular flow is shown in Section 5.1; its impact on the flow  kinetic energy rate is addressed in the Appendix.
\newline

\section{Application to a  Cardiovascular Model}

This Section explores the application of annular flow models to cardiovascular  circulatory systems in the context of heart failure, a degenerative cardiovascular disease where the heart fails to pump enough blood to meet the body's demand. Most heart pumps used to manage heart failure  operate with continuous mechanisms \cite{Slaughter}. 
\newline

Steady-state and oscillatory annular flow analytical solutions derived in the previous Sections highlighted the limiting effect on flow rates caused by the \textit{no-slip} boundary condition on rigid tube surfaces, along with significant velocity reducing viscous effects at the walls. By introducing a coaxial fluid jet, the study shows significant  flow rate increases (up to a factor 2), suggesting potential applications to enhance cardiac output.
\newline

The model under consideration is based on an intra-ventricular small diameter cylindrical  axial pump anchored in the heart's apex,  positioned within the outflow tract of the Left Ventricle (LV), the heart's main pumping chamber, in  proximity to the aortic valve, ejecting a pulsatile blood jet \cite{Garrigue1}. Acting as a flow accelerator, this pump operates synchronously with the heartbeats \cite{Garrigue2}.
\newline

The model exhibits several  limitations which are addressed below. However, it offers analytical tools and valuable insights into optimizing flow performance with pulsatile intra-ventricular or intra-aortic cardiac pumps.
\newline

 \underline{5.1   Numerical Results}
\newline

The data presented in this Section  is derived from the asymptotic annular velocity solutions (\ref{eq:52})  and (54) within  the radial range   $0.25 \le \hat{r} \le 1 $. Figures 3 through 5 illustrate three sets  of   pulsatile  velocity  amplitudes  (n = 1,  n = 5, and the cumulative sum of all harmonic contributions from n = 1 to 5) as functions of the radius $\hat r$ for consecutive, identical phase angles $\omega t$, where $\omega$ is the cardiac frequency.   
\newline

To compute   harmonic velocity amplitudes, as defined in (\ref{eq:4}), the  pulsatile pressure gradient $\widetilde P(t)$ driving the flow is expressed through its Fourier components.  
\newline

Over a cardiac cycle of duration $T$, we consider a simple intra-ventricular pressure  model  $P(t)$ which  consists of two consecutive phases: 

\begin{enumerate}
\item \textbf{Ejection phase}: represented by a  sinusoidal  pulse pressure $\Delta P$ of duration $\tau$ ($0\le t \le \tau$) starting at a high pressure $P_1$. Here $\tau$ denotes the Left Ventricular Ejection Time (LVET).
\item  \textbf{Filling phase}: represented by a  constant low pressure $P_2 \ll P_1$ for the remaining portion of the cardiac cycle.
\end{enumerate}

The pulsatile pressure  profile as a function of time $t$ is thus expressed as
\begin{equation} \label{eq:63}
 \begin{array}{l}
      {P(t) = P_1 + \Delta P \sin{(\pi \frac{t}{\tau})}}  \quad     \text{for}    \quad  0\le t \le \tau \\
      {P(t) = P_2  \quad   \text{for}    \quad  \tau \le t \le T} 
 \end{array}
\end{equation} 

In the case of a pulsatile jet synchronous with the native heart flow, the  velocity profile as a function of time $t$, with $\overline{V}_0$ constant,  is expressed as
\begin{equation} \label{eq:64}
 \begin{array}{l}
      V(t) = \overline{V}_0  \quad     \text{for}    \quad  0\le t \le \tau \\
      V(t)= 0  \quad   \text{for}    \quad  \tau \le t \le T 
 \end{array}
\end{equation} 

The following numerical parameters  are used throughout the analysis (CGS units):

\begin{enumerate}
\item \textbf{Blood Flow}:   $\rho = 1.05$ $g/cm^3$, $\mu = 0.04$  $dyne-s/cm^2$, \cite{Fahraeus}
\item \textbf{Geometry}:   $L $= 10 cm, $\lambda = R_2/R_1 = 4$
\item \textbf{Womersley Parameters}:  $\omega = 8.38$  rad/s, $\alpha_1 = 5$,  $\alpha_2 = 20$
\item \textbf{Pressures}:  $T $= 0.75 s, $\tau$ = 0.23 s, $P_1 = 80$,  $\Delta P = 30$, and $P_2 = 10$ (pressures in units of 1,333.2 $dyne/cm^2$). 
\item \textbf{Jet Velocity}: $\overline{V}_0 = 2.0$
\newline
\end{enumerate}

As outlined  in the discussion on Transverse Wave dynamics, the primary characteristic of  pulsatile velocity profiles (Figs. 3 to 5), is that,  irrespective of the phase angle or  harmonic number, the core  flow exhibits  a flat velocity profile over approximately 50 \% to 70 \% of the annular cross section.
\newline

Figures 3 and 4 depict normalized oscillatory velocity profiles $u_n(\hat{r},t) / u_p$ for harmonics n = 1 and 5, respectively, evaluated at 8 identical phase angle values $\omega t$. The annular flow starts at time t = 0 (label $\omega t_0$)  and ends at $t = \tau$ (label $\omega t_{20}$).
\newline

\begin{figure}[h!]
\centering
\includegraphics[scale=0.41]{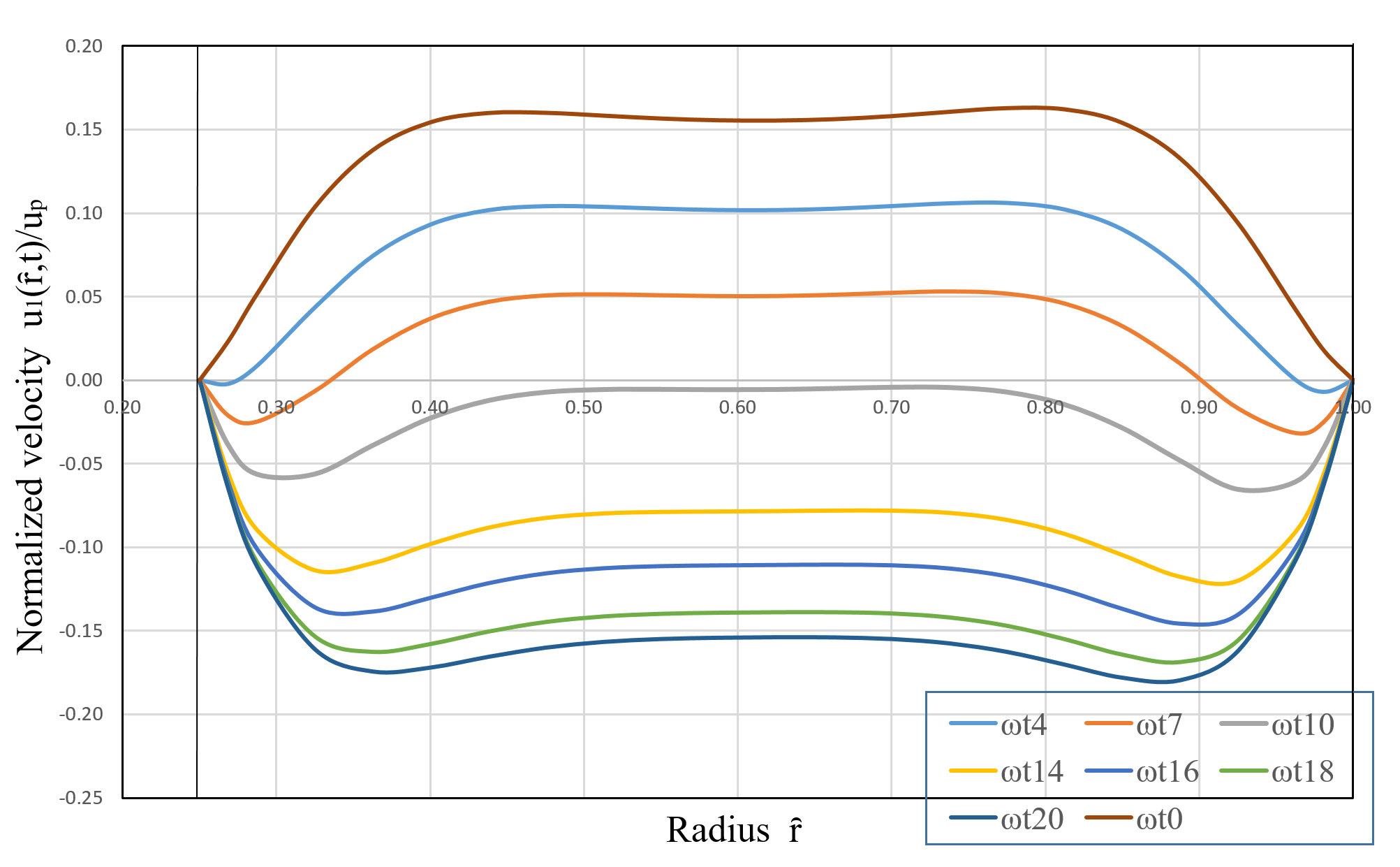}
\caption{Oscillatory annular velocity profiles $u_1(\hat{r},t)/u_p$ as a function of the radius $\hat{r}$ at 8 phase angles;  harmonic n = 1}
\label{fig:Fig3}
\end{figure}

Figure 3 displays the gradual evolution of the oscillatory annular velocity profile as a function of time (from $\omega t_0$ to $\omega t_{20}$) with its flat central core. As predicted by (\ref{eq:52}), near the annular surface boundaries ($\hat{r} \simeq 0.25$ and $ \hat{r} \simeq 1$),  transverse waveforms are generated in viscous layers characterized by strong shear rate variations that may induce velocity reversal. Also, the n = 1 core velocity amplitude  $u_1(\hat{r},t) / u_p$  decreases  from about +0.15 to -0.15: it can be shown to be the opposite for the n = 2 harmonic, $\textit{albeit}$ with  smaller amplitudes.
\newline

In Figure 4, all $5^{th}$-harmonic  core velocity amplitudes $u_5(\hat{r},t) / u_p$ are attenuated by an approximate factor  $|\widetilde{P_1|}/ |\widetilde{P_5}| \simeq  20$ relative  to the n = 1  case. Additionally,  all transverse waveforms near the annular boundaries  exhibit  sharper oscillations confined to increasingly  narrower viscous layers. Also, as expected with higher harmonics, due to rapid phase changes, the core velocity exhibits several sinusoidal-like oscillations over time between the respective -0.007 and +0.007 velocity amplitudes.
\newline

Figure 5 displays the oscillatory annular velocity profiles resulting from the summation of the first five harmonics (n = 1 to 5).  While this summation  does not significantly alter the velocity characteristics  of the first harmonic  (Fig. 3), for the same phase angle,  the core velocity amplitude is reduced down to approximately 70 \% of its initial  n = 1 value, indicating harmonic velocity competition. 

\begin{figure}[h!]
\centering
\includegraphics[scale=0.40]{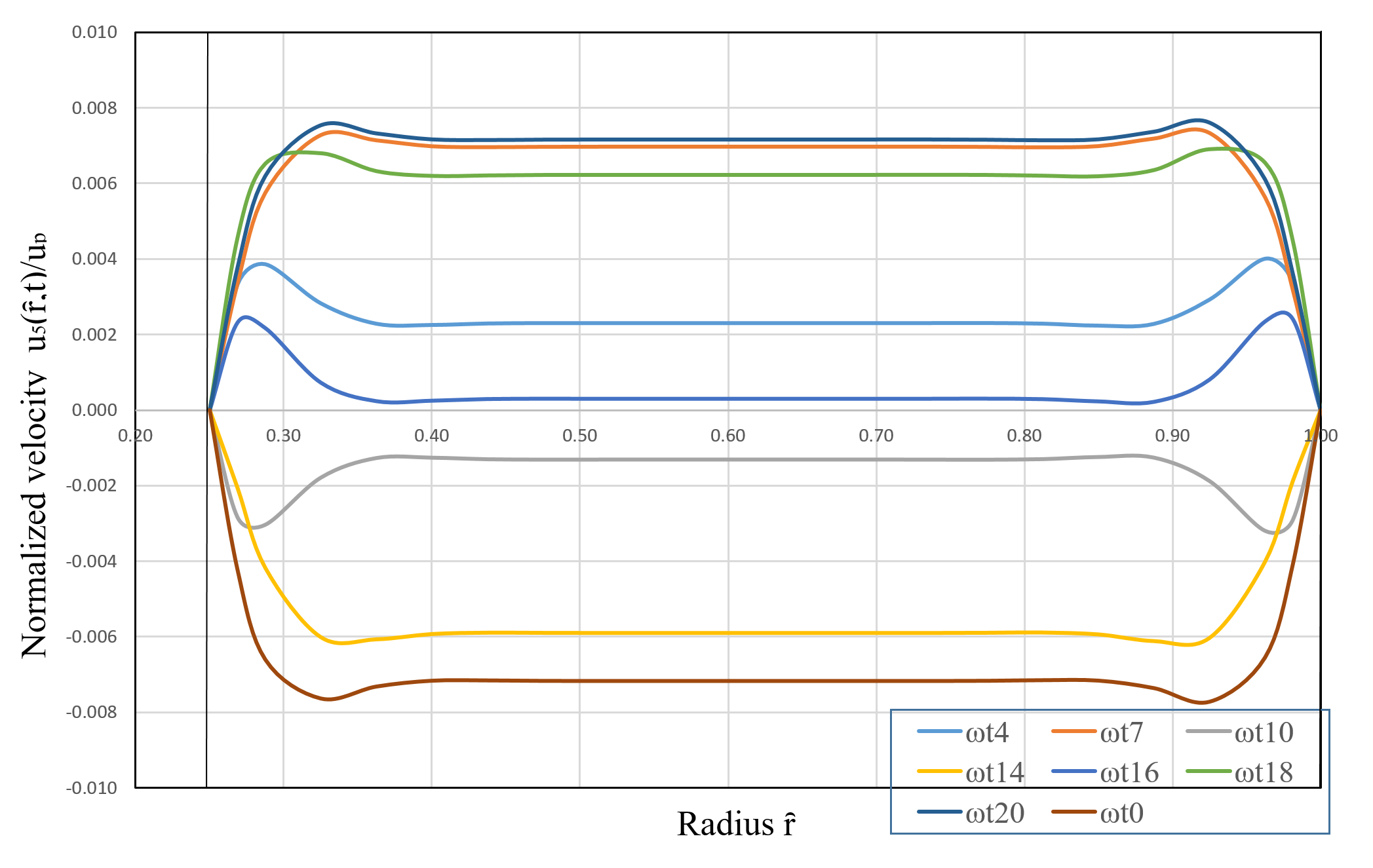}
\caption{Oscillatory annular velocity profiles $u_5(\hat{r},t)/u_p$ as a function of the radius $\hat{r}$ at 8 phase angles;  harmonic n = 5}
\label{fig:Fig4}
\end{figure}

Changes in the viscous waveform near the boundaries, including substantial velocity reversals,  highlight marked  variations in shear-stress as observed by comparing the oscillatory profiles  $\omega t_0$ near the boundaries between Figs. 3 and 5. Also, the annular  velocity profiles are not  symmetrical with respect to the centerline $\hat{r} = 0.625$ as can be observed by comparing the negative magnitudes of the $\omega t_{20}$ profile in the vicinity of $\hat{r} = 0.35$ and $\hat{r} = 0.90$, respectively. 
\newline

\begin{figure}[h!]
\centering
\includegraphics[scale=0.39]{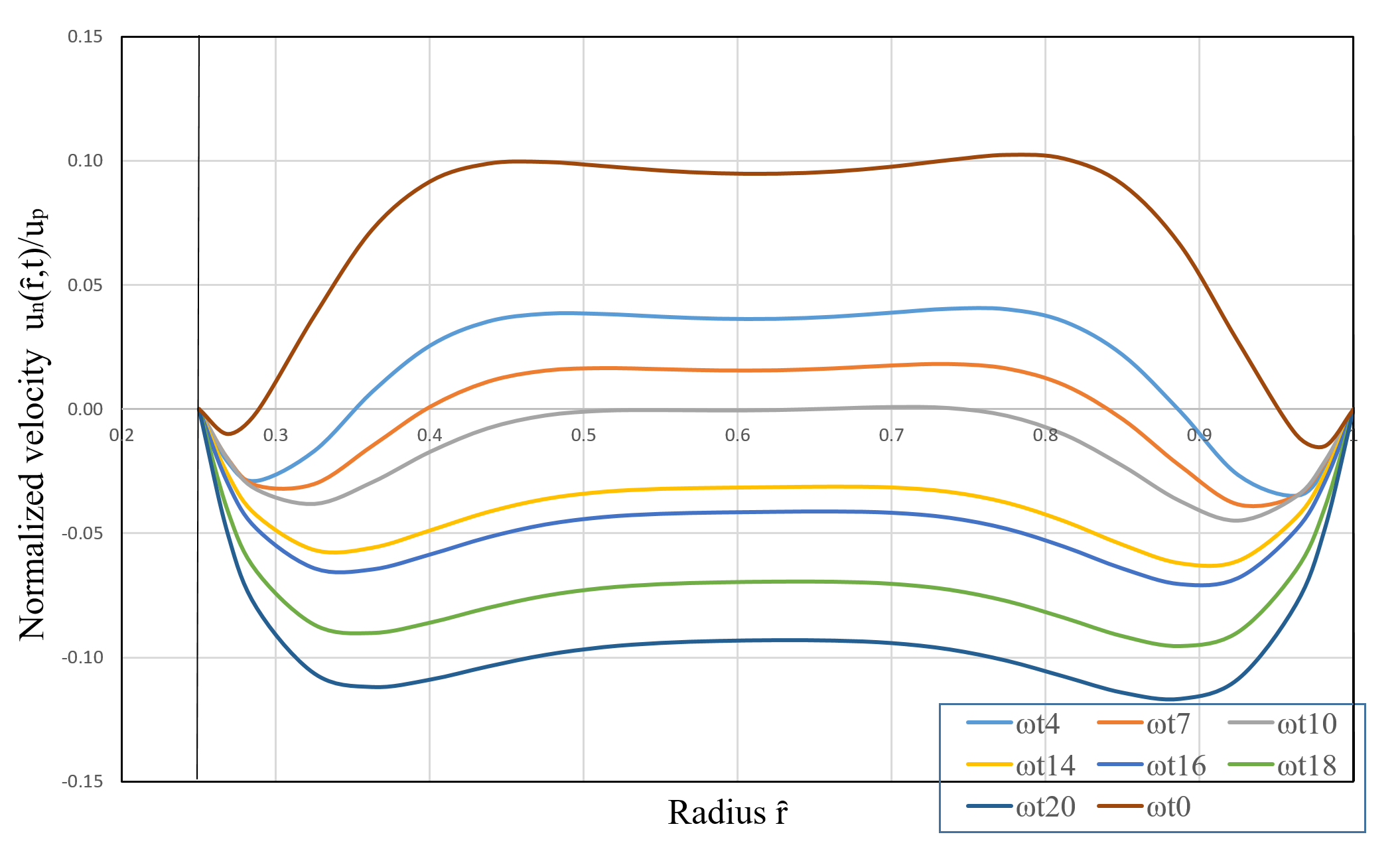}
\caption{Sum of oscillatory annular velocity profiles as a function of the radius $\hat{r}$ at 8 phase angles:  harmonics n = 1 to 5}
\label{fig:Fig5}
\end{figure}

\begin{figure}[h!]
\centering
\includegraphics[scale=0.41]{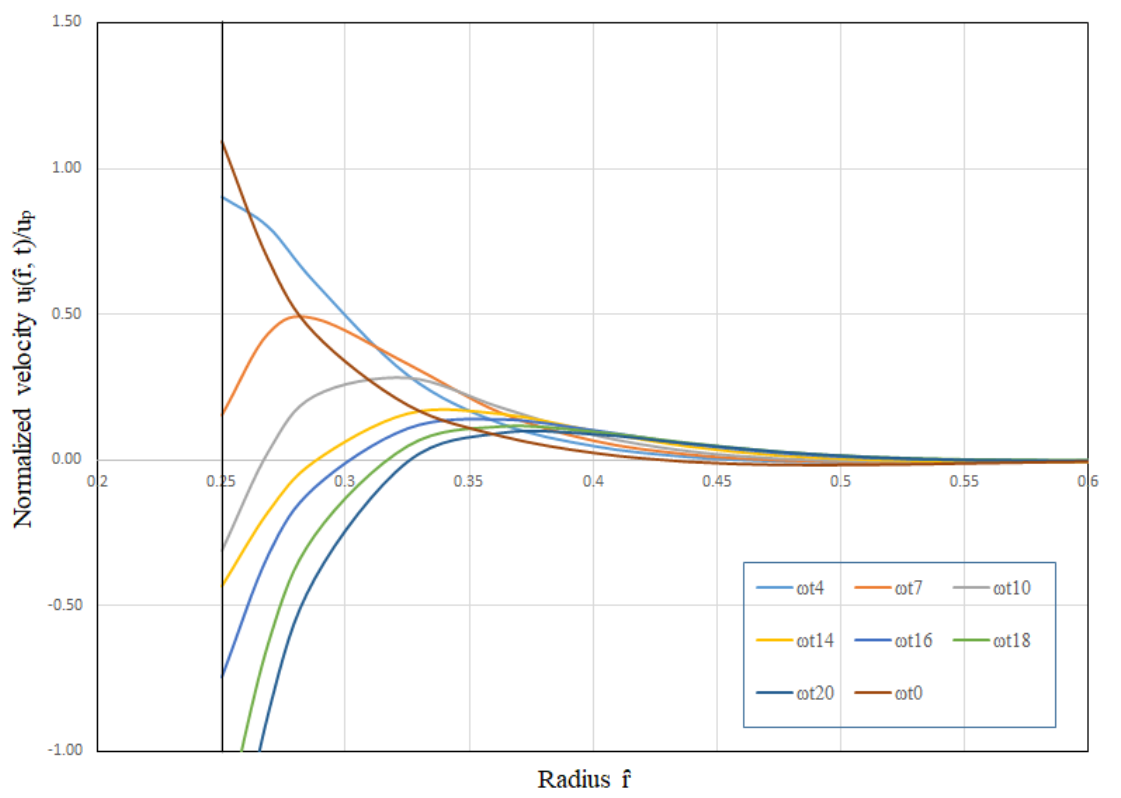}
\caption{Sum of pulsatile jet-induced annular velocity profiles as a function of the radius $\hat{r}$ at 8 phase angles: velocity harmonics n = 1 to 5; $\overline{V}_0 = 2.0$ and $\lambda = 4$. Ejection in the plane $x = 0$.}
\label{fig:Fig6}
\end{figure}

Figure 6 illustrates the jet-induced pulsatile velocity amplitude contribution $u_j(\hat{r},t) / u_p$  presented in  solution (54),  resulting from the summation of the first five annular velocity harmonics (n = 1 to 5). With a constant velocity coaxial jet $(\overline{V}_0 = 2.0)$ ejected  in synchrony with the native heart flow, positive annular velocity  enhancement is achieved at all phase angles sequentially over time, starting with the phase angles $\omega t_0$ to $\omega t_{14}$, gradually  followed by  phase angles $\omega t_{16}$ to $\omega t_{20}$,  with diminishing amplitudes further along the annular radius.
\newline

The $\textit{depth of interaction}$ of the coaxial jet with the pulsatile annular flow is limited to about 45 \% of the annular width, significantly less than in the steady-state case (Fig. 1),  illustrating the strong interaction between pulsatile jet and  annular flow. Superposing the jet velocity $u_j(\hat{r},t)$ to the pulsatile annular velocity  $u_n(\hat{r},t)$ essentially compensates the flow reduction in the vicinity of the inner tube surface caused by the $\textit{no-slip}$ boundary condition, thereby increasing the overall flow-rate. 
\newline

For the case $\overline{V}_0 = 2.0$ and $\lambda = 4$, the approximate jet-induced pulsatile annular flow rate $Q_j(t)$ obtained from (55) for the time period $\omega t = 0$ to  1.93 (end of the jet ejection pulse) is nearly constant and equal to $0.74\; Q_p$.   The total pulsatile annular flow rate $Q_j(t)$, including the steady-state jet induced flow rate $Q_{0,j}$, (\ref{eq:18}), is $Q_j(t) = 3.807 \; Q_p$, thereby increasing the Poiseuille flow rate by a factor of approximately 3.8 under the same conditions.
\newline

\underline{5.2   Limitations of the Model}
\newline

This section addresses key limitations of the model,  in particular  the chosen geometry, the heart wall elasticity, and the fully developed flow assumption.

\begin{enumerate}
\item \underline{Cylindrical Geometry}

Cylindrical models  have been used  to study the human heart ejection performance \cite{Sallin}. The LV geometry considered here is not strictly cylindrical. However, LV blood ejection takes place through the \textit{nearly circular} proximal sub-valvular channel called "Left Ventricle Outflow Tract" (LVOT), facing the aortic valve, where the  axial pump is  positioned.

Despite the complex anatomy of the  LV chamber, in the presence of a cylindrical jet, each ejection results in a pulsatile annular flow propagating along the central cylindrical pump from the heart's apex towards the open aortic valve over approximately 10 cm. The near-field ($x \ge 0$) co-flow takes place in the quasi circular semi-rigid LVOT. Therefore, employing a coaxial cylindrical geometry for  the inner jet together with the simultaneous outer annular flow inside the  LVOT's inlet  serves as a reasonable first approximation for modeling LV  flow dynamics.
\newline

\item \underline{Elastic Heart Walls}

Heart walls are elastic, which  introduces additional complexities into the annular flow model. Periodic cardiac wall contractions, responsible for pulsatile blood ejection, influence velocity profiles, flow rates, and viscous flow dynamics.

To quantify the effect of pulsatile LV  wall  motion on  ejected flow,  the estimated heart muscle wall  radial velocity  is compared to the average intra-ventricular longitudinal  velocity during the LV  Ejection Time. Human systolic time intervals including LVET, have been studied in both healthy individuals as well as in individuals with heart failure: they are shorter in dilated hearts \cite{Weissler}.  The peak ascending aortic velocity is used as a substitute for estimating the axial velocity since the proximal LVOT, the aortic valve, and the ascending aorta have comparable diameters.

In healthy individuals, based on the heart's inner diameter difference between  the  ends of the respective filling  and ejection phases, the average radial wall velocity during ejection is estimated to be in the 4-5 cm/s range and can reach up to 6-7 cm/s in mildly dilated hearts.  These radial wall velocities are less than 5\% of the   $\simeq 120$ cm/s and $\simeq 240$ cm/s respective ascending aorta peak velocities in healthy and dilated hearts   \cite{Stein}.

 Accordingly, a quasi-rigid cylindrical model for the sub-valvular LVOT housing the pulsatile pump which neglects the  small radial flow motion (Sections 3 and 4), can serve as a reasonable $\textit{first approximation}$ for investigating LV annular flow dynamics. A more  realistic pulsatile flow model would account for the  elastic periodic LV wall motion which would superpose a radial velocity component onto the axial annular flow.
\newline




\item \underline{Fully Developed Flow}

The mean peak Reynolds number $R_e$ for    blood flow in the ascending aorta
reaches 5000 or more, indicating turbulent flow \cite{Stein}. For intra-ventricular annular flows under consideration, peak velocities are about 25\% to 30\% of these velocities; hence the corresponding mean peak annular Reynolds number falls in the 1250-1500 laminar range. However, for normal and dilated hearts in the presence of a blood jet in the  proximal LVOT,  $R_e$ might  exceed 2000, indicating that, in this case, the laminar flow assumption might not strictly hold during LV ejection \cite{Shemer}.

 During  this phase, the axial pump  recirculates viscous blood from the  apical region  towards the aortic annulus over a distance  approximately 5 times the average annular gap  before mixing with the pump jet. 
In this  $x \ge 0$ region,  the turbulent pulsatile jet flow propagates  an additional 1 to 2 cm  into the LVOT, ensuring an effective mixing between the jet  and the entrained   native flow. In  steady-state conditions, this entrainment  extends far into the annular space, as illustrated in Figs. 1 and 2, whereas in pulsatile conditions it extends to approximately 30\%  of the annular width. 

From these considerations, in particular the short propagation distance of the annular flow, the  assumption of fully developed pulsatile co-flow established within the LVOT might warrant further caution \cite{Shemer}.


\end{enumerate}

Despite these limitations, the idealized rigid cylindrical coaxial geometry offers a reasonable benchmark for analyzing intra-ventricular oscillatory annular flows. However, incorporating elastic heart wall motion into future studies will be essential to develop a more realistic model of intra-cardiac pulsatile  blood flows. 

\section{Conclusion}
This study provides exact analytical  solutions for both steady-state and oscillatory annular viscous flows in coaxial cylindrical systems. Furthermore, practical asymptotic solutions expressed in terms of \textit{real functions} are derived for pulsatile flows in annular geometries. These solutions demonstrate the critical interplay of viscous and inertial forces, including the presence of transverse waves near  boundary surfaces, as well as flow direction reversals. The introduction of a coaxial fluid jet significantly enhances the velocity profiles and flow rates across the annular section. 
\newline

The findings of this study hold practical implications for cardiovascular applications, such as optimization of intra-ventricular heart pumps. Despite some limitations, the analytical model presented  here offers valuable insights into optimizing annular flow performance and addressing challenges associated with induced pulsatile cardiac flows. Future work will focus on extending this model to incorporate elastic wall effects, which are crucial for capturing the dynamics of more realistic cardiovascular systems.

\section{Appendix : Energy Equation}
The kinetic energy rate of increase  $\dot{E}_{KE}$ of an incompressible fluid enclosed in a volume V is
\begin{equation} \label{eq:60}
\dot{E}_{KE} = \frac{\partial }{\partial t} \int_{v} (\rho \frac{u^2}{2}) dV =  \int_{v} \rho u  \frac{\partial u}{\partial t} dV              
\end{equation}

Inserting the equation of motion (\ref{eq:2}) into (60) together with the annular volume element   $dV = L dA$ and cross-section $dA =  2 \pi r dr$ yields
\begin{equation} \label{eq:61}
\dot{E}_{KE} = - L\frac{d}{dx} \int_{R_1}^{R_2} p u dA + 2\pi L \mu (r u \frac{\partial u}{\partial r})|_{R_1}^{R_2}- \mu L \int_{R_1}^{R_2} (\frac{\partial u}{\partial r})^2 dA
\end{equation}

The first integral represents the work done by the pressure forces per unit time; the last integral, always $\ge 0$, represents the rate of energy dissipation in the incompressible fluid  due to   viscous forces,  which always diminish the mechanical energy.
\newline

In the absence of a jet, the middle term vanishes by virtue of the \textit{no-slip} boundary condition (\ref{eq:3}). Noticeably $\mu \frac{\partial u}{\partial r}$ is the x-component of the viscous stress tensor $\sigma_{xr}$ defined in (\ref{eq:56}) which is finite on these surfaces. 
\newline

In this oscillatory no-jet case, the kinetic energy rate $\dot{E}_{KE}$ in (\ref{eq:61}) expresses the instantaneous balance between the energy flux of the pressure forces in the x-direction and the rate of energy dissipation through the annular cross-section in which the  pulsatile velocity and annular  flow rate are  given by (\ref{eq:31}) and (\ref{eq:36})
\begin{equation} \label{eq:62}
\dot{E}_{KE} = - \frac{L}{\rho}\frac{dp}{dx} Q(t) - \mu L \int_{R_1}^{R_2} (\frac{\partial u}{\partial r})^2 dA
\end{equation}

In the steady-state, since energy is conserved the kinetic energy rate $\dot{E}_{KE} = 0$ .  This is verified   by inserting  $u_0(r)$ from (\ref{eq:12}) together with  $Q_0$ from (\ref{eq:14}) into (\ref{eq:62}).
\newline

In the presence of a coaxial jet of instantaneous velocity $V_0(t)$ ejected in the plane $x = 0$, the applicable boundary condition at the surface of radius $R_1$ for the induced secondary stream is $u(R_1) = V_0(t)$. This introduces a pulsatile surface term in (\ref{eq:61}) which, as seen in Section 4.2.3 can either be positive or negative, yielding
\begin{equation} \label{eq:63}
\dot{E}_{KE, j} = - \frac{L}{\rho}\frac{dp}{dx} Q(t) - 2\pi R_1 L V_0(t) \mu  \frac{\partial u}{\partial r}|_{R_1}   - \mu L \int_{R_1}^{R_2} (\frac{\partial u}{\partial r})^2 dA
\end{equation}

In the steady-state,  the first and last terms in (\ref{eq:63}) cancel out: the rate $\dot{E}_{KE,j}$ reduces to a surface term expressing the entrainment of the outer annular flow by the coaxial pulsatile velocity jet. When $\overline{V}_0 \ge 1$, the surface term is always positive.




\nocite{}

\bibliographystyle{abbrv}


\bibliographystyle{plain}
\end{document}